%% file: main.tex
\documentclass[lettersize,journal]{IEEEtran}
\usepackage{amsmath,amsfonts}
\usepackage{algorithmic}
\usepackage{algorithm}
\usepackage{array}
\usepackage[caption=false,font=normalsize,labelfont=sf,textfont=sf]{subfig}
\usepackage{textcomp}
\usepackage{stfloats}
\usepackage{url}
\usepackage{verbatim}
\usepackage{graphicx}
\usepackage{cite}
\DeclareMathOperator{\E}{E}

\usepackage[colorlinks=false]{hyperref}
\DeclareMathOperator*{\argmax}{argmax}

\begin{document}

\title{Sparse Regression Codes exploit Multi-User Diversity without CSI}

\author{
\IEEEauthorblockN{VSV Sandeep, Sai Dinesh Kancharana, Arun Pachai Kannu.}
      \thanks{This work is supported in part by the Qualcomm 6G UR grant and in part by the Prime Minister's Research Fellowship (PMRF).
      The authors are with the Department of Electrical Engineering, Indian Institute of Technology Madras, Chennai - 600036, Tamil Nadu, India. (email: ee22d004@smail.iitm.ac.in, ee20d401@smail.iitm.ac.in, arunpachai@ee.iitm.ac.in).
      }
}

\maketitle

\begin{abstract}
We study sparse regression codes (SPARC) for multiple access channels with multiple receive antennas, in non-coherent flat fading channels. 
We propose a novel practical decoder, referred to as maximum likelihood matching pursuit (MLMP), which greedily finds the support of the codewords of users with partial maximum likelihood metrics. As opposed to the conventional successive-cancellation based greedy algorithms, MLMP works as a successive-combining energy detector. We also propose MLMP modifications to improve the performance at high code rates. Our studies in short block lengths show that, even without any channel state information, SPARC with MLMP decoder achieves multi-user diversity in some scenarios, giving better error performance with multiple users than that of the corresponding single-user case. We also show that SPARC with MLMP performs better than conventional sparse recovery algorithms and pilot-aided transmissions with polar codes.   
\end{abstract}


\section{Introduction}
Hyper/ultra-reliable and low-latency communication (HRLLC/URLLC) in 6G/5G \cite{wang2023_6G} targets mission-critical applications with the fundamental requirement of reliable communication in short block lengths \cite{durisi2016_SBC}. This is a challenging problem, since the error performance of coding schemes gets worse as the block length decreases \cite{shannon1959probability}. 

Sparse regression codes (SPARC) \cite{joseph2012least} are error control codes where transmitted codewords are generated from sparse linear combinations of columns from a dictionary matrix. SPARC are shown to be asymptotically capacity-achieving under approximate message passing decoding \cite{rush2017,barbier2017} in additive white Gaussian noise (AWGN) channels. SPARC have been studied in the short block lengths regime for URLLC applications in \cite{ji2018svc,ji2019PL-SVC,Madhu_APK}. SPARC for multi-user channels with perfect channel state information (CSI) are considered in \cite{ji2018svc,SVC_MU_MIMO21}, and a successive interference cancellation based matching pursuit is developed in \cite{SVC_MU_MIMO21}. Non-coherent SIMO channels are investigated in \cite{ji2019PL-SVC} using block orthogonal matching pursuit at the receiver. SPARC for the grant-free multiple access channel has been studied for the asymptotic block lengths in \cite{amalladinne2020coded,fengler2021sparc}.

In our work, we consider the multiple access fading channel where the channel fading coefficients are unknown at both the transmitter and the receiver. The conventional techniques employ pilot symbols for channel estimation \cite{tongpilot}, the base station utilises the CSI feedback for allocating the resources among the users, and then the data transmission phase starts, resulting in a large latency. We consider SPARC in non-coherent multi-user channels with multiple receive antennas and derive novel decoders which does not need CSI. We present a non-coherent maximum likelihood (ML) detector for SPARC over MIMO flat-fading channels. Since optimal ML decoding is computationally infeasible due to exhaustive search requirements, we propose maximum likelihood matching pursuit (MLMP), a practical greedy algorithm that iteratively selects the active columns in the SPARC codeword based on partial ML metrics. Our MLMP decoder works based on successive combining principles as opposed to the conventional successive cancellation techniques. We introduce two enhanced variants to MLMP: MLMP with replacement and Parallel MLMP in order to improve the error performance at higher code rates. 

Using simulation results, we show that the proposed SPARC with the MLMP decoder performs better than the pilot-aided transmissions (PAT) employing polar codes with successive cancellation list decoding assisted by cyclic redundancy check \cite{balatsoukas2015polarllr}. We also show the superior performance of the proposed MLMP decoder over the conventional sparse signal recovery decoders such as orthogonal matching pursuit (OMP) \cite{tropp_omp}, and compressive sampling matching pursuit (CoSaMP) \cite{needell_cosamp}. Our results also demonstrate that SPARC with non-coherent MLMP decoders enjoy multi-user diversity gains for high code rates, that is, the error performance is better when $3$-users are sending their bits than the single user scenario (while the overall spectral efficiency is the same). 

\section{SPARC encoder and decoder}
\subsection{SPARC Encoding for multi-user case}
SPARC encoder employs a dictionary matrix $\mathbf{A}$ of size $N \times L$, which is partitioned into $K$ sections such that $\mathbf{A} = \bigl[\mathbf{A}_1 \, \mathbf{A}_2 \, ... \, \mathbf{A}_K\bigr]$  with section $\mathbf{A}_k$ having $L_k$ number of columns. We construct a complex-valued dictionary matrix using mutually unbiased bases from quantum information theory \cite{WOOTTERS,Madhu_APK}, which has $L=N^2$ columns and the maximum pairwise inner product is $\sqrt{1/N}$. In a multi-user channel with $U$ users (with $U\leq K$), the $K$ sections in the dictionary are allocated among $U$ users so that each user gets a unique set. Let $\mathcal{A}_j = \{\mathbf{A}_{j,1} \, \mathbf{A}_{j,2} \, ... \, \mathbf{A}_{j,K_j}\} \subset \{\mathbf{A}_1 \, \mathbf{A}_2 \, ... \, \mathbf{A}_K \}$ represents the ordered set of $K_j$ sections allocated to user-$j$, with $\mathcal{A}_j \cap \mathcal{A}_p = \emptyset$ and $\sum_{j = 1}^{U} K_j = K$. Based on its information bits, user-$j$ selects one column from each section and obtains the codeword as  
\begin{equation}
    \mathbf{s_j} = \sum_{k = 1}^{K_j} \mathbf{a}_{j,k}, \label{codeword}
\end{equation}
where $\mathbf{a}_{j,k}$ is the column chosen from the section $\mathbf{A}_{j,k}$. Since we consider non-coherent channels, we use unmodulated SPARC \cite{rush2017}, where the chosen columns are directly added without multiplication by any constellation symbols. With $L_{j,k}$ being the number of columns in section  $\mathbf{A}_{j,k}$, the number of information bits encoded by user-$j$ is $N_{b_j} = \sum_{k=1}^{K_j} \log_2 L_{j,k}$, and the total number of bits transmitted by all the users is $\sum_{j=1}^{U} N_{b_j} = N_b$. In this encoding process, using a total of $2N$ real dimensions, we convey a total of $N_b$ bits from $U$ users and denote this as $(2N,N_b,U)$ coding scheme, with the code rate $\frac{N_b}{2N}$.

\subsection{Received signal model}
Our system model comprises a multiple access channel with multiple user equipment (UE), each with a single transmit antenna, communicating with a base station (BS) having $M$ receive antennas. The channel between user-$j$ and receive antenna $i$ of the BS is modeled by an independent Rayleigh flat-fading channel coefficient $h_{i,j} \sim \mathcal{CN}(0,\sigma_{j}^2)$ (circularly-symmetric complex Gaussian distribution with mean $0$ and variance $\sigma_j^2$). The received signal for a $U$-user scenario at the $i^{th}$ receive antenna is given by
\begin{equation}
    \mathbf{y_i} = \sum_{j=1}^U h_{i,j} \mathbf{s}_j +\mathbf{v}_i, \quad \forall i \in [M], \label{ML_eq1}
\end{equation}
where $[M]$ denotes the set $\{1,2,\ldots,M\}$, with white Gaussian noise $\mathbf{v}_i \sim \mathcal{CN}(0,\sigma_{v}^2 \mathbf{I}_N)$. We consider the non-coherent scenario where the channel coefficients $h_{i,j}$ are unknown at the receiver. We assume that the variances of both the channel fading coefficients and the noise are known at the receiver.
 
\subsection{Maximum Likelihood Detector}
SPARC decoding involves recovering the \emph{active} columns, which contributed to the construction of the codeword $\mathbf{s_j}$ (\ref{codeword}), for all the users $j \in [U]$.  We first derive the ML detector for this model and then present our proposed MLMP greedy detector.
With $h_{i,j} \sim \mathcal{CN}(0,\sigma_{j}^2)$ and $\mathbf{v}_i \sim \mathcal{CN}(0,\sigma_{v}^2 \mathbf{I}_N)$, the probability distribution of the observation $\mathbf{y}_i$ conditioned on the transmitted codewords is
\begin{equation}
    \mathbf{y}_i|\mathbf{s}_1, \mathbf{s}_2,..., \mathbf{s}_U \sim \mathcal{CN}(0,\underbrace{\sum_{j=1}^U \sigma_{j}^2 \mathbf{s}_j \mathbf{s}_j^*+ \sigma_{v}^2 \mathbf{I}_N}_{\mathbf{F}}),  \label{ML_eq2}
\end{equation}
where, $(.)^*$ indicates conjugate transpose.
Since the fading coefficients are independent across the antennas, the ML metric (MLM) with $M$ receive antennas is obtained as
\begin{align}
&\hspace{-12mm}\mathrm{MLM}(\mathbf{y}_1,...,\mathbf{y}_M,\mathbf{s}_1, \cdots, \mathbf{s}_U,\sigma_v^2) \nonumber \\ 
& \quad \quad = \sum_{i=1}^M \log p\left(\mathbf{y}_i \mid \mathbf{s}_1, \cdots, \mathbf{s}_U, \sigma_{v}^2\right), \label{ML_eq5}\\ 
& \quad \quad = -M\log \left|\mathbf{F}\right| -  \sum_{i=1}^M\mathbf{y}_i^* {\mathbf{F}}^{-1} \mathbf{y}_i. \label{ML_eq6}
\end{align}
The ML detector output is 
\begin{align}
    \hat{\mathbf{s}}_1, \cdots, \hat{\mathbf{s}}_U & =  \underset{\mathbf{s}_1, \cdots, \mathbf{s}_U}{\arg \max} \left[ -M\log \left|\mathbf{F}\right| -  \sum_{i=1}^M\mathbf{y}_i^* {\mathbf{F}}^{-1} \mathbf{y}_i \right]. \label{ML_eq6_2}
\end{align}

In general, the ML metric computation involves inverting the covariance matrix $\mathbf{F}$ in (\ref{ML_eq2}). However, due to the special structure of $\mathbf{F}$, we can use the Woodbury identity to compute the inverse. For a 3-user case, the ML metric term $\mathbf{y_i}^* {\mathbf{F}}^{-1} \mathbf{y_i}$ is given in (\ref{ML_eq10}), where $\gamma_3$ is given by (\ref{ML_eq11}). Also, the determinant  $\left|\mathbf{F}\right|$ evaluates to  $\left|\mathbf{F}\right| = (\lambda_1 + \sigma_{v}^2)(\lambda_2 + \sigma_{v}^2)(\lambda_3 +\sigma_{v}^2)(\sigma_{v}^2)^{N-3}$ with $\lambda_1, \lambda_2, \lambda_3$ being eigenvalues of $3 \times 3$ matrix $\mathbf{B^{*} B}$ where $\mathbf{B} = [{\sigma_{1}} \mathbf{s_1} \, \, {\sigma_{2}} \mathbf{s_2} \, \, {\sigma_{3}} \mathbf{s_3}]$. For a 2-user case, we can obtain the ML metrics by setting $\mathbf{s}_3=\mathbf{0}$ and for a single user case by setting $\mathbf{s}_1=\mathbf{0}$ and $\mathbf{s}_2=\mathbf{0}$, in the 3-user metric expressions. A detailed analysis of the single-user case is made in \cite{kancharana2024sparse}.
\begin{figure*}[t]
\begin{align}
\small
& -\mathbf{y_i}^* {\mathbf{F}}^{-1} \mathbf{y_i}\overset{(3)}{=} 
\frac{1}{\sigma_{v}^4} \frac{1}{\gamma_3} 
\left\{ \left( \left[\frac{1}{\sigma_{2}^2}+\frac{\left\|\mathbf{s_2}\right\|^2} {\sigma_{v}^2}\right] \left[\frac{1}{\sigma_{3}^2}+\frac{\left\|\mathbf{s_3}\right\|^2}
{\sigma_{v}^2}\right] - \frac{\left|\left\langle \mathbf{s_2, s_3}\right\rangle\right|^2}
{\sigma_{v}^4} \right)
\left|\left\langle \mathbf{s_1, y_i}\right\rangle\right|^2 
+ \left(
  \left[\frac{1}{\sigma_{1}^2}+\frac{\left\|\mathbf{s_1}\right\|^2}
 {\sigma_{v}^2}\right] 
 \left[\frac{1}{\sigma_{3}^2}+\frac{\left\|\mathbf{s_3}\right\|^2}
{\sigma_{v}^2}\right]  \right. \right. \nonumber \\ 
& \left. \left. - \frac{\left|\left\langle \mathbf{s_1, s_3}\right\rangle\right|^2} {\sigma_{v}^4} \right) \left|\left\langle \mathbf{s_2, y_i}\right\rangle\right|^2 + \left( \left[\frac{1}{\sigma_{1}^2}+\frac{\left\|\mathbf{s_1}\right\|^2}{\sigma_{v}^2}\right] \left[\frac{1}{\sigma_{2}^2}+\frac{\left\|\mathbf{s_2}\right\|^2}{\sigma_{v}^2}\right]   - \frac{\left|\left\langle \mathbf{s_1, s_2}\right\rangle\right|^2} {\sigma_{v}^4} \right) \left|\left\langle \mathbf{s_3, y_i}\right\rangle\right|^2 -2 \operatorname{Re} \left[ \left(  \left[\frac{1}{\sigma_{3}^2}+\frac{\left\|\mathbf{s_3}\right\|^2} {\sigma_{v}^2}\right] \frac{\left\langle \mathbf{s_2, s_1}\right\rangle}{\sigma_{v}^2} \right. \right.\right.  \nonumber \\
& \left. -\frac{\left\langle \mathbf{s_2, s_3}\right\rangle\left\langle \mathbf{s_3, s_1}\right\rangle}{\sigma_{v}^4}\right) \left\langle \mathbf{y_i, s_2}\right\rangle\left\langle \mathbf{s_1, y_i}\right\rangle + \left( \left[\frac{1}{\sigma_{2}^2}+\frac{\left\|\mathbf{s_2}\right\|^2} {\sigma_{v}^2}\right] \frac{\left\langle \mathbf{s_3, s_1}\right\rangle}{\sigma_{v}^2} -\frac{\left\langle \mathbf{s_2, s_1}\right\rangle\left\langle \mathbf{s_3, s_2}\right\rangle}{\sigma_{v}^4} \right) \left\langle \mathbf{y_i, s_3}\right\rangle\left\langle \mathbf{s_1, y_i}\right\rangle + \left( \left[\frac{1}{\sigma_{1}^2}+\frac{\left\|\mathbf{s_1}\right\|^2} {\sigma_{v}^2}\right] \right. \nonumber\\
&\left.  \left. \frac{\left\langle \mathbf{s_3, s_2}\right\rangle}{\sigma_{v}^2} - \frac{\left\langle \mathbf{s_1, s_2}\right\rangle\left\langle \mathbf{s_3, s_1}\right\rangle}{\sigma_{v}^4} \right) \left\langle \mathbf{y_i, s_3}\right\rangle\left\langle \mathbf{s_2, y_i}\right\rangle \right\}. \label{ML_eq10}\\
& \gamma_3 = \left(\frac{1}{\sigma_1^2}+\frac{\left\|\mathbf{s_1}\right\|^2}{\sigma_{v}^2}\right)
\left(\frac{1}{\sigma_2^2}+\frac{\left\|\mathbf{s_2}\right\|^2}{\sigma_{v}^2}\right)
\left(\frac{1}{\sigma_3^2}+\frac{\left\|\mathbf{s_3}\right\|^2}{\sigma_{v}^2}\right)
- \left(\frac{1}{\sigma_1^2}+\frac{\left\|\mathbf{s_1}\right\|^2}{\sigma_{v}^2}\right)
\frac{\left|\left\langle \mathbf{s_2, s_3}\right\rangle\right|^2}
 {\sigma_{v}^4}
 - \left(\frac{1}{\sigma_3^2}+\frac{\left\|\mathbf{s_3}\right\|^2}{\sigma_{v}^2}\right)
\frac{\left|\left\langle \mathbf{s_1, s_2}\right\rangle\right|^2}
 {\sigma_{v}^4} \nonumber \\&
 - \left(\frac{1}{\sigma_2^2}+\frac{\left\|\mathbf{s_2}\right\|^2}{\sigma_{v}^2}\right)
\frac{\left|\left\langle \mathbf{s_1, s_3}\right\rangle\right|^2}
 {\sigma_{v}^4}
  +2 \operatorname{Re}\left\{\frac{1}{\sigma_{v}^6} \left\langle \mathbf{s_1, s_2}\right\rangle\left\langle \mathbf{s_2, s_3}\right\rangle\left\langle \mathbf{s_3, s_1}\right\rangle\right\} \label{ML_eq11}. 
\end{align}
\hrulefill
\vspace*{-0.2cm}
\end{figure*}

From (\ref{ML_eq10}), the ML detection metric depends on the energy of the correlations between the observation $\mathbf{y}_i$ and the user codewords $\mathbf{s}_1, \mathbf{s}_2$, and $\mathbf{s}_3$. The metric also depends on the correlations between the codewords of different users $\langle \mathbf{s}_j, \mathbf{s}_k \rangle$. 
ML detector in (\ref{ML_eq6_2}) searches over all the possible codewords of all the users jointly, and this search complexity grows exponentially with the number of sections $K$. As a special case, when the codewords are mutually orthogonal, which corresponds to orthogonal resource allocation (i.e., $\mathbf{s}_j^* \mathbf{s}_k = 0$ for $j \neq k$), 
the ML metric in (\ref{ML_eq6_2}) decouples as
$\displaystyle -\sum_{j=1}^{U} (M \log(\sigma_j^2 \|\mathbf{s}_j\|^2 + \sigma_{v}^2) + \frac{1}{\sigma_{v}^4} \frac{1}{\left(\frac{1}{\sigma_{j}^2}+\frac{\left\|\mathbf{s}_j\right\|^2}{\sigma_{v}^2}\right)} \sum_{i=1}^M \left|\left\langle \mathbf{s}_j, \mathbf{y}_i\right\rangle\right|^2) .
$
With orthogonal codewords, the maximization of the ML metric decouples across users, enabling independent decoding for each user.

\subsection{Maximum Likelihood Matching Pursuit}
We propose Maximum Likelihood Matching Pursuit (MLMP), a greedy iterative algorithm that identifies one active column per iteration using partial ML metrics, while treating contributions from undetected columns as noise. The top column identified in an iteration can be from any user. Since the algorithm identifies only one active column in each iteration, the search complexity grows linearly with the number of sections $K$.

\textbf{Definitions:} 
The section $A_k$ in the dictionary is referred to as identified if the algorithm has already found the active column in that section. Let $\mathcal{A}_j^{(t)}$ denote the set of all identified sections corresponding to the user-$j$ until the $t^{th}$ iteration, and the number of identified sections is $K_j^{(t)}$. Let the partially detected codeword of user-$j$, $\hat{\mathbf{s}}_j^{(t)}$, be the sum of all active columns from the identified sections of user-$j$ until the $t^{th}$ iteration. These quantities are initialized as $\hat{\mathbf{s}}_j^{(0)} = \mathbf{0}$, $\mathcal{A}_j^{(0)} = \emptyset, \, \forall j \in [U].$

\textbf{Metric Computation:}
In the $(t+1)$-th iteration, we find an active column from the unidentified sections as follows. We compute the partial ML metric for each of the candidate (hypothesis) column, by assuming that all the identified columns until the current iteration are correct and the remaining yet to be identified columns act as white Gaussian noise. Defining $\tilde{\mathbf{s}}_j^{(t)} = \mathbf{s}_{j} -  \hat{\mathbf{s}}_j^{(t)}$, we have
\begin{equation}
\mathbf{y_i} = \sum_{j=1}^U h_{i,j} \hat{\mathbf{s}}_j^{(t)} + \sum_{j=1}^U h_{i,j} \tilde{\mathbf{s}}_j^{(t)} + \mathbf{v_i}. \label{mlmp_eq2}
\end{equation}
Variance of the noise due to the unidentified columns is obtained by dividing the average energy by the block length as
\begin{align}
 \sigma_{u}^2 &= \frac{1}{N} \E \{ \|  \sum_{j=1}^U h_{i,j} \tilde{\mathbf{s}}_j^{(t)} \|^2\}. 
\end{align}
Since the channel fading coefficients are independent, we have $\sigma_{u}^2 = \frac{1}{N}   \sum_{j=1}^U \sigma_j^2 \E \{ \| \tilde{\mathbf{s}}_j^{(t)} \|^2\}$. Under the mild assumption that $\E\{ \langle \mathbf{a}_i, \mathbf{a}_j \rangle \} = 0$ for randomly chosen columns $\mathbf{a}_i$ and $\mathbf{a}_j$ from the dictionary, we have $\E \{ \| \tilde{\mathbf{s}}_j^{(t)} \|^2\} = K_j - K_j^{(t)}$, which is the number of unidentified sections of user-$j$ until the $t^{th}$ iteration.

Now, we calculate the partial ML metric for a candidate (hypothesis) column $\mathbf{a}_h$ for user-$j$ from one of its undetected sections $\mathbf{a}_h \in \mathcal{A}_j \setminus \mathcal{A}_j^{(t)}$. While doing that, we change the codeword hypothesis for the $j^{th}$ user as $\hat{\mathbf{s}}_j^{(t)}+\mathbf{a}_h$ and retain the partial codewords of other users as such. Since the candidate column gets added to the codeword hypothesis, the effective noise (including the contributions from the remaining undetected columns) is $\sigma_{e,j}^2 = \sigma_v^2 + \sigma_u^2 - \frac{1}{N}\sigma_j^2$. Specifically, the partial ML metric is obtained in the form given by (\ref{ML_eq6}) by passing the arguments as $\mathrm{MLM}(\mathbf{y}_1,\cdots, \mathbf{y}_M, \hat{\mathbf{s}}_1^{(t)}, \cdots, \hat{\mathbf{s}}_j^{(t)}+\mathbf{a}_h, \cdots, \hat{\mathbf{s}}_U^{(t)}, \sigma_{e,j}^2)$.

\textbf{Finding the top column:} Now, among all the undetected columns from all the users, we select the top column in the $(t+1)^{th}$ iteration with the highest partial ML metric as 
\begin{multline*}
   (\hat{j},\hat{\mathbf{a}}_h)
   = \argmax_{{{\mathbf{a}}_h \in \mathcal{A}_j \setminus \mathcal{A}_j^{(t)}, \,j \in [U]}} \mathrm{MLM}(\mathbf{y}_1,\cdots,\mathbf{y}_M, \hat{\mathbf{s}}_1^{(t)}, \cdots , \\ \hat{\mathbf{s}}_j^{(t)}+\mathbf{a}_h, \cdots, \hat{\mathbf{s}}_U^{(t)} , \sigma_{e,j}^2   ).
\end{multline*}
The algorithm \emph{combines} each candidate column $\mathbf{a}_h$ with already detected partial codewords $\hat{\mathbf{s}}_j^{(t)}$ and selects the one with the highest (partial) ML metric.  

\textbf{Update the parameters:} Supposing that the top column found above $\hat{\mathbf{a}}_h$ from the user-$\hat{j}$ belongs to the section $\mathbf{A}_k$, the parameters are updated as $\hat{\mathbf{s}}_{\hat{j}}^{(t+1)} = \hat{\mathbf{s}}_{\hat{j}}^{(t)} +\hat{\mathbf{a}}_h $ and $\mathcal{A}_{\hat{j}}^{(t+1)} = \mathcal{A}_{\hat{j}}^{(t)} \cup \mathbf{A}_k$. The rest of the users' partial codewords and the corresponding detected sections are retained as such.

\textbf{Stopping condition:} The algorithm terminates after $K$ iterations when complete support recovery is achieved. 

\subsection{MLMP Extensions}
In greedy algorithms, the first selected column is error-prone due to interference from columns of undetected sections (\ref{mlmp_eq2}). Early selection of inactive columns propagates errors through subsequent iterations \cite{kwon_shim_mmp}. To address this, we introduce MLMP with replacement (MLMP-R) and parallel-MLMP (P-MLMP) as enhancements to MLMP.

MLMP-R operates in two phases. Phase 1 runs standard MLMP for $K$ iterations to find total support. Phase 2 runs additional $K$ iterations for revalidation. During revalidation, for each top column from a section, the algorithm reevaluates the ML metric across all columns in that section while keeping other sections' top columns fixed. If another column from that section yields a higher ML metric, it replaces the previously selected top column; otherwise, the previously selected top column is confirmed for that section. This two-phase approach fine-tunes the initially detected support.

Parallel MLMP selects $P$ top columns with the highest ML metrics in the first iteration. Treating each as a probable active column, $P$ parallel MLMP instances are run simultaneously. After $K$ iterations, each instance produces a separate codeword with its detected columns, yielding $P$ estimated codewords. Among these $P$ candidates, the final codeword is selected as the one with the highest ML metric value.

\subsection{Computational complexity}

To compare the computational complexity of sparse recovery algorithms, we measure the number of candidate hypotheses for which the ML detection metric is computed in order to recover the complete support $K$. Table \ref{table1} lists the exact and approximate complexity expressions in terms of the number of sections $K$ and the size of the dictionary $L$. As an example, we have given the comparison for $K = 4$ and $L = 256$ columns. MLMP requires the fewest detection metric evaluations, followed by MLMP-R, OMP and CoSaMP. Parallel MLMP with 8 paths is computationally expensive, and the complexity increases linearly with the number of parallel paths. Ideal ML decoding with exhaustive search is computationally infeasible.

\begin{table*}[t]
\centering
\caption{Comparison of computational complexity}
\label{table1}
\begin{tabular}{|c|c|c|c|}
\hline
\multicolumn{1}{|c|}{\textbf{Decoding algorithm}} & \multicolumn{1}{c|}{\textbf{Complexity(Exact)}}  & \multicolumn{1}{c|}{\textbf{Complexity (Approxmn)}} & \multicolumn{1}{c|}{\textbf{In numbers ($K = 4, L = 256$)}}\\ [0.5mm] \hline
 Ideal ML & $(L/K)^K$ & $(L/K)^K$ & 16777216 \\ \hline
 OMP & $KL - K(K-1)/2$ & $KL$ & 1018\\ \hline
 CoSaMP & $2KL -2K^2 + K$ & $2KL$ & 2020\\ \hline
 MLMP & $(K+1) L/2$ & $(K+1) L/2$ & 640\\ \hline
 MLMP-R & $(K+3) L/2$ & $(K+3) L/2$ & 896\\ \hline
 Parallel-MLMP ($P$ paths) & $P (K-1) L/2 + L$ & $P (K-1) L/2 + L$ & 3328 ($P=8$)\\ \hline
\end{tabular}
\end{table*}

\subsection{Pilot-Aided Transmissions with Polar Codes}
For non-coherent channels, pilot-aided transmission (PAT) \cite{apkannu2008pat,kannu2005mse} are typically employed where pilot symbols are sent to facilitate the channel estimation and the estimated channel is used for coherently decoding the data. For comparison with SPARC, we consider a PAT scheme coupled with polar codes for transmitting data. Each user is assigned dedicated resources so that polar codes do not get any inter-user interference. For the multi-user SPARC scheme with parameters $(2N,N_b,U)$, each user transmits approximately $\frac{2N}{U}$ bits. Hence, for the PAT scheme, each user employs $(\frac{2N}{U}, \frac{N_b}{U})$ polar code so that the total number of resources used and the total number of bits transmitted match with the SPARC scheme.   

\section{Simulation results}
We evaluate our system performance using block error rate (BLER). We declare successful decoding only when \emph{all} the users' information bits are correctly decoded.
We plot BLER versus $E_b/N_0$ (in dB) with receiver noise variance  $N_0 = \sigma_v^2$. From (\ref{ML_eq1}), the expected energy per bit is $E_b =  \sum_{j=1}^u \sigma_j^2 \frac{K_j}{N_b}$. When the fading channel $h_{i,j}$ variances $\sigma_j^2$ are identical across users and all the users getting approximately equal number of bits, the individual users' $E_b$ is same as the overall $E_b$. We set the channel fading variances as $\frac{1}{M}$ so that the total received signal power remains the same, regardless of the number of receive antennas. 

\begin{figure}[!t]
\centering
\includegraphics[width=\columnwidth, height=2.2in]{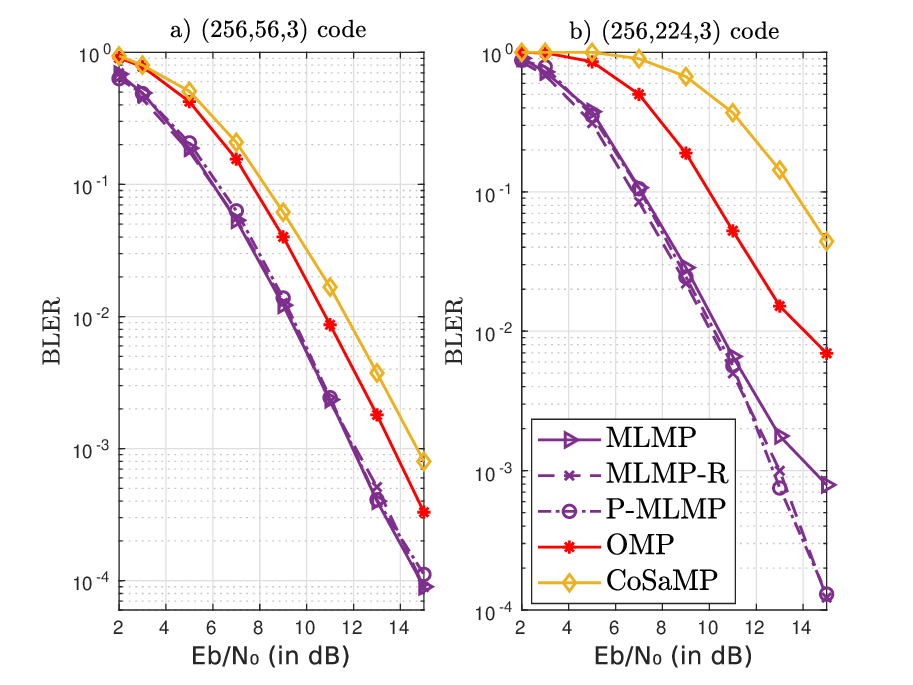}
\caption{Comparison with existing sparse signal recovery algorithms.}
\label{figure_1}
\end{figure}

\begin{figure}[!t]
\centering
\includegraphics[width=\columnwidth, height=2.2in]{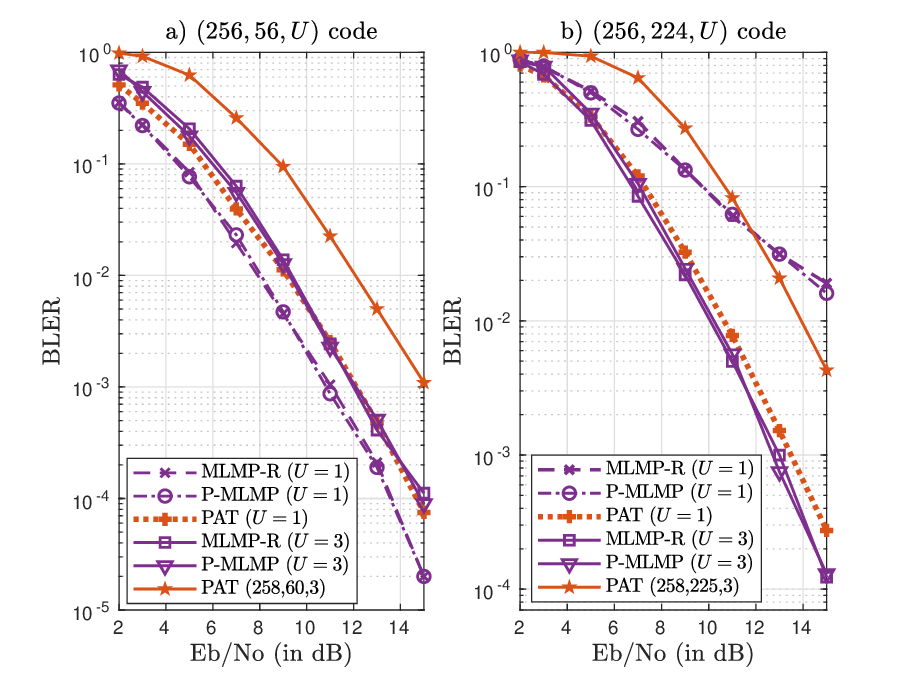}
\caption{Comparison with PAT polar codes.}
\label{figure_2}
\end{figure}

We use the MUB dictionary with  $N = 128$ ($L = N^2 = 128^2$ columns) and $M=4$ receive antennas. We consider symmetric channels ($\sigma_1^2 = \sigma_2^2 = ... = \sigma_u^2$), and uniform section allocation ($K_j = \frac{K}{U}$) for Figures \ref{figure_1} and \ref{figure_2}. For Parallel MLMP results, we consider $P=8$ paths. For Polar PAT, we follow the 5G NR uplink standard with CRC length 11 and successive cancellation list decoder, with list size 8.

Figure \ref{figure_1}(a) compares BLER performance for SPARC scheme (256,56,3) with $K=6$ sections and \ref{figure_1}(b) shows the results for (256,224,3) scheme with $K=24$ sections. MLMP outperforms conventional sparse recovery algorithms like OMP and CoSAMP, as it uses partial ML metrics and successive combining for selecting the active columns. Also, when the code rates ($\frac{N_b}{2N}$) are high ($K$ is high),    
the MLMP enhancements (MLMP-R and P-MLMP) show notable improvements over MLMP. 

Figure \ref{figure_2} compares BLER performance for single-user ($U=1$) and three-user ($U=3$) systems for SPARC and polar PAT at lower-rate $(256,56,U)$ (Figure \ref{figure_2}(a)) and higher-rate $(256,224,U)$ (Figure \ref{figure_2}(b)). At the lower rate, single-user systems perform slightly better than three-user systems for both SPARC and polar PAT. However, at the higher rate, three-user SPARC achieves multi-user diversity gain by exploiting independent fading across users, where some users have better channel conditions than others. Though the MLMP decoder does not know the channel fading coefficients, the ML metrics will naturally be higher for the columns corresponding to the strong users, and hence MLMP ends up recovering the support of strong users first. After that, MLMP faces a simpler task of decoding the remaining support (which will be a smaller set) of the other users. On the other hand, PAT schemes allocate a fraction of energy to pilot symbols, and the energy available for data transmission will be correspondingly reduced. PAT schemes also suffer from channel estimation errors. In addition, due to the allocation of dedicated resources between users, each user transmits information bits in a reduced block length $\frac{2N}{U}$. It is well known that for the same code rate, when the block length decreases, the error probability increases \cite{shannon1959probability}. Hence, polar PAT performs well for the single-user case but performs poorly for the $3$-user case. Our proposed SPARC decoder has about $4$ dB gain over the PAT polar scheme for the 3-user case. 

\begin{figure}[!t]
\centering
\includegraphics[width=\columnwidth, height=2in]{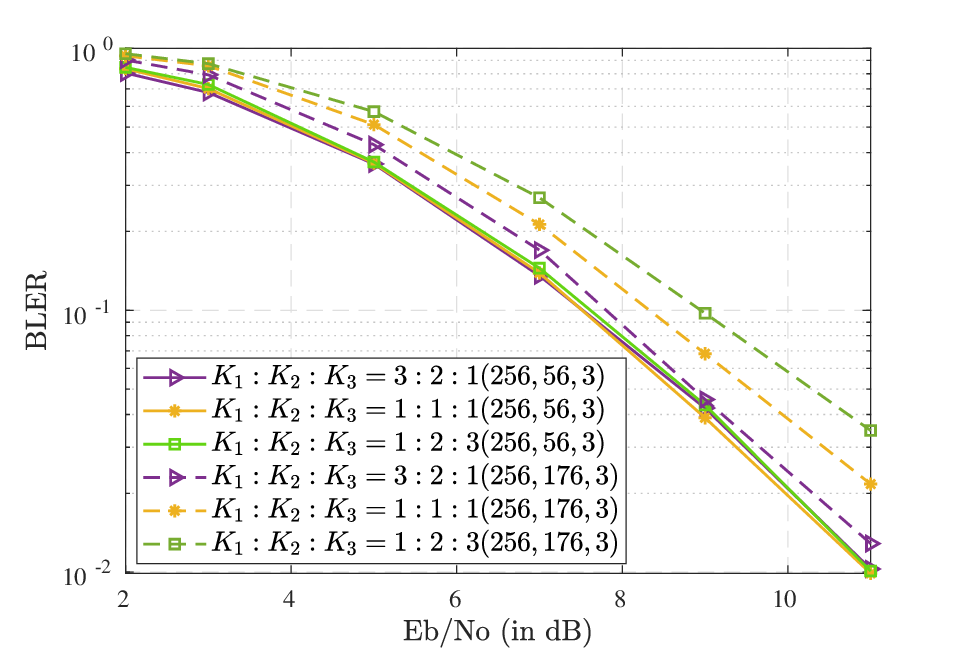}
\caption{Asymmetric fading conditions with $\sigma_1^2:\sigma_2^2:\sigma_3^2 = 3:2:1$. }
\label{figure_3}
\end{figure}

In practical scenarios with asymmetric channel conditions, rate adaptation strategies are crucial for maximizing BLER performance. Figure \ref{figure_3} investigates rate adaptation for a 3-user system with channel fading variances $\sigma_1^2:\sigma_2^2:\sigma_3^2 = 3:2:1$ using MLMP. Here $E_b$ refers to the overall energy per bit including all 3 users. Since the received signal power is proportional to channel variance, user 1 has the highest probability of successful support recovery, followed by users 2 and 3. We compare three section allocation schemes: (i) proportional to channel strengths ($K_1:K_2:K_3 = 3:2:1$), (ii) reverse proportional ($K_1:K_2:K_3 = 1:2:3$), and (iii) equal allocation ($K_1:K_2:K_3 = 1:1:1$). At lower code rates ($(256,56,3)$ code with $K=6$), all schemes achieve similar BLER performance. However, at higher rates ($(256,176,3)$ code with $K=18$), proportional allocation yields the best performance, followed by equal allocation, while reverse proportional allocation performs poorly.

\section{Conclusion}

This work explored SPARC for multi-user multiple access MIMO systems in fading channels with non-coherent detection for short block lengths. We proposed the MLMP algorithm, which employs a partial maximum likelihood metric for support recovery and demonstrates superior block error rate performance over conventional algorithms. We further introduced MLMP-R and parallel MLMP as enhancements that achieve better performance than MLMP at higher code rates. Our simulation results highlight SPARC's advantage in exploiting multi-user diversity, particularly at higher code rates, without requiring CSI, offering significant benefits over traditional methods such as pilot-aided transmissions with polar codes.

\input{main.bbl}

\end{document}

%% file: main.bbl

%% file: main.bbl
\begin{thebibliography}{10}
\providecommand{\url}[1]{#1}
\csname url@samestyle\endcsname
\providecommand{\newblock}{\relax}
\providecommand{\bibinfo}[2]{#2}
\providecommand{\BIBentrySTDinterwordspacing}{\spaceskip=0pt\relax}
\providecommand{\BIBentryALTinterwordstretchfactor}{4}
\providecommand{\BIBentryALTinterwordspacing}{\spaceskip=\fontdimen2\font plus
\BIBentryALTinterwordstretchfactor\fontdimen3\font minus \fontdimen4\font\relax}
\providecommand{\BIBforeignlanguage}[2]{{%
\expandafter\ifx\csname l@#1\endcsname\relax
\typeout{** WARNING: IEEEtran.bst: No hyphenation pattern has been}%
\typeout{** loaded for the language `#1'. Using the pattern for}%
\typeout{** the default language instead.}%
\else
\language=\csname l@#1\endcsname
\fi
#2}}
\providecommand{\BIBdecl}{\relax}
\BIBdecl

\bibitem{wang2023_6G}
C.-X. Wang, X.~You, X.~Gao, X.~Zhu, Z.~Li, C.~Zhang, H.~Wang, Y.~Huang, Y.~Chen, H.~Haas, J.~S. Thompson, E.~G. Larsson, M.~D. Renzo, W.~Tong, P.~Zhu, X.~Shen, H.~V. Poor, and L.~Hanzo, ``On the {R}oad to {6G}: {V}isions, {R}equirements, {K}ey {T}echnologies, and {T}estbeds,'' \emph{IEEE Communications Surveys \& Tutorials}, vol.~25, no.~2, pp. 905--974, 2023.

\bibitem{durisi2016_SBC}
G.~Durisi, T.~Koch, and P.~Popovski, ``Toward {M}assive, {U}ltrareliable, and {L}ow-{L}atency {W}ireless {C}ommunication {W}ith {S}hort {P}ackets,'' \emph{Proceedings of the IEEE}, vol. 104, no.~9, pp. 1711--1726, 2016.

\bibitem{shannon1959probability}
C.~E. Shannon, ``Probability of error for optimal codes in a {G}aussian channel,'' \emph{Bell System Technical Journal}, vol.~38, no.~3, pp. 611--656, 1959.

\bibitem{joseph2012least}
A.~Joseph and A.~R. Barron, ``Least {S}quares {S}uperposition {C}odes of {M}oderate {D}ictionary {S}ize {A}re {R}eliable at {R}ates up to {C}apacity,'' \emph{IEEE Transactions on Information Theory}, vol.~58, no.~5, pp. 2541--2557, 2012.

\bibitem{rush2017}
C.~Rush, A.~Greig, and R.~Venkataramanan, ``Capacity-{A}chieving {S}parse {S}uperposition {C}odes via {A}pproximate {M}essage {P}assing {D}ecoding,'' \emph{IEEE Transactions on Information Theory}, vol.~63, no.~3, pp. 1476--1500, 2017.

\bibitem{barbier2017}
J.~Barbier and F.~Krzakala, ``Approximate {M}essage-{P}assing {D}ecoder and {C}apacity {A}chieving {S}parse {S}uperposition {C}odes,'' \emph{IEEE Transactions on Information Theory}, vol.~63, no.~8, pp. 4894--4927, 2017.

\bibitem{ji2018svc}
H.~Ji, S.~Park, and B.~Shim, ``{Sparse Vector Coding for Ultra Reliable and Low Latency Communications},'' \emph{IEEE Transactions on Wireless Communications}, vol.~17, no.~10, pp. 6693--6706, 2018.

\bibitem{ji2019PL-SVC}
H.~Ji, W.~Kim, and B.~Shim, ``{Pilot-Less Sparse Vector Coding for Short Packet Transmission},'' \emph{IEEE Wireless Communications Letters}, vol.~8, no.~4, pp. 1036--1039, 2019.

\bibitem{Madhu_APK}
M.~K. Sinha and A.~P. Kannu, ``{Generalized Sparse Regression Codes for Short Block Lengths},'' \emph{IEEE Transactions on Communications}, vol.~72, no.~5, pp. 2536--2551, 2024.

\bibitem{SVC_MU_MIMO21}
R.~Zhang, B.~Shim, Y.~Lou, S.~Jia, and W.~Wu, ``Sparse {V}ector {C}oding {A}ided {U}ltra-{R}eliable and {L}ow-{L}atency {C}ommunications in {M}ulti-{U}ser {M}assive {MIMO S}ystems,'' \emph{IEEE Transactions on Vehicular Technology}, vol.~70, no.~1, pp. 1019--1024, 2021.

\bibitem{amalladinne2020coded}
V.~K. Amalladinne, J.-F. Chamberland, and K.~R. Narayanan, ``{A Coded Compressed Sensing Scheme for Unsourced Multiple Access},'' \emph{IEEE Transactions on Information Theory}, vol.~66, no.~10, pp. 6509--6533, 2020.

\bibitem{fengler2021sparc}
A.~Fengler, P.~Jung, and G.~Caire, ``{SPARCs for Unsourced Random Access},'' \emph{IEEE Transactions on Information Theory}, vol.~67, no.~10, pp. 6894--6915, 2021.

\bibitem{tongpilot}
L.~Tong, B.~Sadler, and M.~Dong, ``Pilot-assisted wireless transmissions: general model, design criteria, and signal processing,'' \emph{IEEE Signal Processing Magazine}, vol.~21, no.~6, pp. 12--25, 2004.

\bibitem{balatsoukas2015polarllr}
A.~Balatsoukas-Stimming, M.~B. Parizi, and A.~Burg, ``{LLR-Based Successive Cancellation List Decoding of Polar Codes},'' \emph{IEEE Transactions on Signal Processing}, vol.~63, no.~19, pp. 5165--5179, 2015.

\bibitem{tropp_omp}
J.~A. Tropp and A.~C. Gilbert, ``{Signal Recovery From Random Measurements Via Orthogonal Matching Pursuit},'' \emph{IEEE Transactions on Information Theory}, vol.~53, no.~12, pp. 4655--4666, 2007.

\bibitem{needell_cosamp}
D.~Needell and J.~Tropp, ``{CoSaMP: Iterative signal recovery from incomplete and inaccurate samples},'' \emph{Applied and Computational Harmonic Analysis}, vol.~26, no.~3, pp. 301--321, 2009.

\bibitem{WOOTTERS}
W.~K. Wootters and B.~D. Fields, ``Optimal state-determination by mutually unbiased measurements,'' \emph{Annals of Physics}, vol. 191, no.~2, pp. 363--381, 1989.

\bibitem{kancharana2024sparse}
S.~D. Kancharana, M.~K. Sinha, and A.~P. Kannu, ``{Sparse Regression Codes for Non-Coherent SIMO channels},'' \emph{arXiv preprint arXiv:2405.09915}, 2024.

\bibitem{kwon_shim_mmp}
S.~Kwon, J.~Wang, and B.~Shim, ``{Multipath Matching Pursuit},'' \emph{IEEE Transactions on Information Theory}, vol.~60, no.~5, pp. 2986--3001, 2014.

\bibitem{apkannu2008pat}
A.~P. Kannu and P.~Schniter, ``{Design and Analysis of {MMSE} Pilot-Aided Cyclic-Prefixed Block Transmissions for Doubly Selective Channels},'' \emph{IEEE Transactions on Signal Processing}, vol.~56, no.~3, pp. 1148--1160, 2008.

\bibitem{kannu2005mse}
A.~Kannu and P.~Schniter, ``{MSE}-optimal training for linear time-varying channels,'' in \emph{Proceedings. (ICASSP '05). IEEE International Conference on Acoustics, Speech, and Signal Processing, 2005.}, vol.~3, 2005, pp. iii/789--iii/792 Vol. 3.

\end{thebibliography}
